\documentclass[preprint,aps,prb,superscriptaddress]{revtex4-1}
\usepackage{amsmath}
\usepackage[pdftex]{graphicx}
\usepackage{ulem}
\usepackage{color}

\begin{document}
\title{Control of the ionization state of 3 single donor atoms in silicon. }

\author{B. Voisin}
\affiliation{SPSMS, UMR-E CEA / UJF-Grenoble 1, INAC, 17 rue des Martyrs, 38054 Grenoble, France}
\author{M. Cobian}
\affiliation{SPMM, UMR-E CEA / UJF-Grenoble 1, INAC, 17 rue des Martyrs, 38054 Grenoble, France}
\author{X. Jehl}
\affiliation{SPSMS, UMR-E CEA / UJF-Grenoble 1, INAC, 17 rue des Martyrs, 38054 Grenoble, France}
\author{M. Vinet}
\affiliation{CEA, LETI, MINATEC Campus, 17 rue des Martyrs, 38054 Grenoble, France}
\author{Y.-M. Niquet}
\affiliation{SPMM, UMR-E CEA / UJF-Grenoble 1, INAC, 17 rue des Martyrs, 38054 Grenoble, France}
\author{C. Delerue}
\affiliation{IEMN, 41 boulevard Vauban, 59046 Lille, France}
\author{S. de Franceschi}
\affiliation{SPSMS, UMR-E CEA / UJF-Grenoble 1, INAC, 17 rue des Martyrs, 38054 Grenoble, France}
\author{M. Sanquer}
\email[]{marc.sanquer@cea.fr}
\affiliation{SPSMS, UMR-E CEA / UJF-Grenoble 1, INAC, 17 rue des Martyrs, 38054 Grenoble, France}

\date{\today}

\begin{abstract}

By varying the gate and substrate voltage in a short silicon-on-insulator trigate field effect transistor we control the ionization state of three arsenic donors. We obtain a good quantitative agreement between 3D electrostatic  simulation and experiment for the control voltage at which the ionization takes place. It allows us observing the three doubly occupied states As$^-$ at strong electric field in the presence of nearby source-drain electrodes. 

\end{abstract}

\maketitle






Although doping has always been the cornerstone of semiconductor technology, devices have started to enter a new era where a single dopant can be used for new (quantum) functionalities: charge and spin qubits, single-electron pumps, turnstiles and transistors~\cite{Zwanenburg2013,Roche2013}.
  
Thin silicon-on-insulator (SOI) devices are particularly attractive for the implementation of donor-based functionalities, because they offer very good control of the transverse electric field in the channel. This electrostatic property is at the core of the Metal-Oxide-Semiconductor Field-Effect Transistors (MOSFETs) but is also crucial to address dopants individually and to control their electronic wavefunctions and couplings. These abilities are prerequisite for dopant-based applications.


In this work we study both experimentally and with simulations how 3 arsenic donors are charged in a nanoscopic MOSFET when the substrate, gate and drain voltages are varied with respect to the grounded source. The ionization state of each donor is separately read out by detecting its corresponding resonance in the source-drain current. The ionization state of each donor - As$^+$, As$^0$ and As$^-$- is individually controlled at low temperature by the electric field. The scalability of a compact system of a few tunable shallow donors beyond the previously studied cases of 1~\cite{Pierre2009,Khalafalla2007,Khalafalla2009,Lansbergen2008,Fuechsle2012} and 2 donors~\cite{Khalafalla2007,Roche2012a} is then shown.

\begin{figure}[t]
\begin{center}
\includegraphics[width=8.5cm]{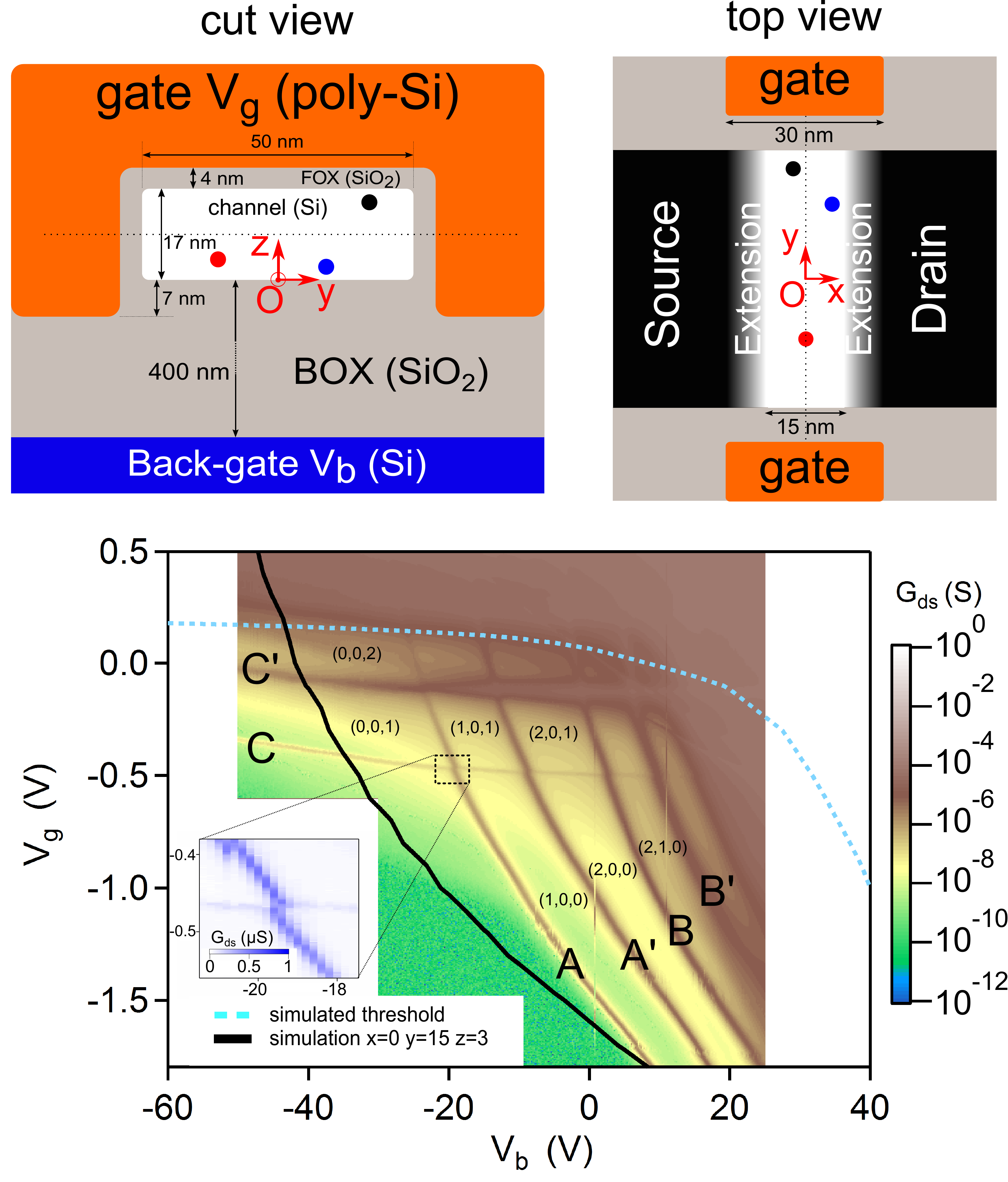}
\caption{(color online). Top panel: Sample layouts. Bottom panel: Color plot of the source-drain conductance versus $V_{b}$ and $V_{g}$ at T=4.2\,K. The dashed line is the simulated threshold voltage of a device without donors. 6 ionization lines noted A to C' are observed to come by pairs. This is attributed to the $As^+$/$As^0$ and $As^0$/$As^-$ ionization lines resulting from the double occupation of 3 different donors. The ionization state (a,b,c) for the three donors is indicated between the lines. States with more than 4 electrons in the body are barely defined because of the strong coupling with the conduction band. According to our simulation the donor corresponding to lines (C,C') is closer to the front gate than the two donors given the "A" and "B" ionization lines. The black line is the ionization line simulated for a donor located at (x,y,z)=(0\,nm, 15\,nm, 3\,nm), see text. Inset: Crossing of the A and C ionization lines at T=1\,K showing a weak Coulomb repulsion.}
\label{fig1}
\end{center}
\end{figure}

In our small MOSFETs -in the 10 nm range size- dopants are not isolated in the channel but see a complex electrostatic environment which includes other donors in the channel and in the source-drain (S-D) as well as offset charges in the gate stack. This environment should be considered cautiously. First other donors in the channel may result in the many-body problem of a Coulomb glass.  For lightly doped semiconductors, long range fluctuation effects dominate over the immediate environment charges~\cite{Efros1975,Efros1984}. Fortunately in our short MOSFETs the interaction between donors is screened by the S-D. We can therefore treat the charging of a specific donor taking the ionization state of the others as constant over a large range of gate voltages, and therefore assign each line in the stability diagram to a specific dopant atom (see Fig.\ref{fig1}).

The ionization of donors at the graded edges -or extensions- of the source and drain is also explicitly considered in our simulation  in a mean-field approach which neglects Kondo~\cite{Lansbergen2010} and Fermi edge singularity ~\cite{Matveev1992} effects, which are not observed in our devices  at 4.2\,K.
The simulation of a realistic electrostatic environment explains the evolution of the ionization lines as function of the control voltages (the front gate voltage $V_g$ and the substrate bias  voltage $V_b$).




The samples, fabricated on 200\,mm SOI wafers, are similar to those described in ref.~\cite{Pierre2009}. A 200\,nm long, 17\,nm thick and 50-nm-wide silicon nanowire was etched from the SOI film and covered at its centre by a 30\,nm long polysilicon gate isolated by a 4\,nm-thick SiO$_2$ layer, called the front oxide (FOX) (see top panels of Fig.~\ref{fig1}). This top gate covers three sides of the silicon channel. A 400\,nm thick buried oxide (BOX) separates the channel from the silicon substrate, which can be biased using the procedure described in ref.~\onlinecite{Roche2012}. 
The central part of the channel contains a few arsenic donors as estimated by process simulations including the rapid thermal annealing step for donors activation~\cite{Pierre2009}.
Conductance was measured at $T$=4.2\,K with a lock-in technique using an a.c. voltage (100 to 300\,$\mu$V) added to the d.c. S-D bias. Radio-frequency filtering was achieved with lossy coaxial lines.
  
The extensions of the S-D are located below the gate because there is no spacers. Therefore the channel length (between 10 and 20~nm) is significantly smaller than the nominal gate length (30\,nm)~\cite{Pierre2009}. Thanks to this small length the donors centered in the channel have sufficient tunnel coupling to both source and drain for their ionization state to be detected by resonant tunneling.

  
The bottom panel of Fig. \ref{fig1} shows the S-D conductance as function of $V_b$ and $V_g$  without d.c. bias. 
Above a certain threshold voltage the S-D current has contributions coming from the conduction band states, while donors in the body of the SOI contribute below this threshold.

The conduction band states are diffusive states that contribute to a continuum of drain current. Donors can be detected by their resonant tunneling contribution to the S-D current when their energy states lie between the Fermi energies of the S-D~\cite{Pierre2009}.

At $V_b$=0 the front channel conduction starts at $V_{th} \simeq $+0.0V/0.1\,V. The onset of the conduction band is indeed expected at $V_{g} = V_{th} \simeq \Delta \phi_{i} + {E_g \over 2 \vert e \vert} \simeq $+0.0\,V for an n$^{++}$ silicon gate at $T$=0\,K (where $\Delta \phi_{i} = -{E_g \over 2 \vert e \vert}$ is the work function difference between the gate electrode and the intrinsic channel) \cite{Taur2000}. $V_b \simeq V_g \simeq $ 0 is the flat-band regime without evidence for donor states.

At $V_b \gg$ 0 the conduction band edge appears at negative values of $V_g$. A large electric field is then present in the channel. The positive value of the substrate bias is balanced by the negative value of the gate voltage. This is also obtained in our simulation (see later on and dashed line in  Fig.~\ref{fig1}). The carriers are accumulated at the BOX interface and the threshold shows a cusp in the ($V_g$, $V_b$) plane (see Fig.~\ref{fig1}). This cusp has already been described in the case of P-doped, macroscopic SOI films ~\cite{Ono2006}.
In that case the cusp is due to the ionization of donors in the body of the channel when a vertical electric field is applied. In the middle of a long undoped channel the cusp is due to the shift of the 2DEG from the top to the backdate interface \cite{Verduijn2013}. It is different in our short nanoscopic FETs in which the S-D electrodes end up with a strong gradient of As atoms near the channel.  In that case
the density of carriers and the ionization state of the donors in the extensions of the S-D can change with varying front and back gate voltages.  In  short devices, this affects the potential landscape in the whole channel and the curvature of the threshold line~\cite{supplemental}. The response of the donors and electron gas in these extensions therefore controls both the slope of the threshold voltage and of the ionization lines of the isolated donors inside the channel (see later on). 





We have simulated the ionization lines of a few donors in the channel of these trigate transistors. For that purpose, we have treated the few impurities in the channel as interacting point charges, and the impurities in the highly-doped source and drain extensions as a continuum. A small, 7\,nm deep overetch of the BOX was taken into account in the simulation (see Fig 1. top panels).

We have first computed the potential landscape $V(\vec{r})$ in the nanowire channel at zero S-D bias~\cite{supplemental}. To this aim, we have solved Poisson's equation self-consistently using the Fermi integral $F_{1/2}( \frac {E_c-eV(\vec{r})-\mu} {k_{B}T})$ as an approximation for the local density of electrons. The density of ionized impurities in the S-D extensions is approximated using Ref.~\onlinecite{Altermatt2006}. Here $E_c$ and $\mu$ are respectively the conduction band edge of silicon and the chemical potential in the device. The $\sim 15$ nm long channel was left undoped, and the simulations were run at 30\,K for computational reasons. This simple model shall give a fair account of the screening by the quasi-metallic source and drain extensions. It provides, admittedly, a coarse description of the channel, but our interest here is the physics of individual impurities, thus below the channel threshold.

Once the potential landscape has been computed as a function of $V_b$ and $V_g$, we have added a few bulk-like impurities in the channel at positions $\vec{r}_i$, and have tracked their bound-state energy levels $E_{1s}(V_b, V_g, \vec{r}_i)=E_c-E_b-eV(\vec{r}_i)$ (where $E_b$ is the binding energy of these impurities, 53\,meV). We have also computed the Coulomb interactions $U_{ij}$ between these impurities (as the screened Coulomb interactions between point charges). We have finally used these data as input for a Coulomb-blockade-like model of the system of impurities, in order to determine the ionization lines of each donor.

First of all, we have computed the expected threshold voltage as the line where the electron concentration (integrated over the thickness of the SOI) exceeds $10^{11}$\,cm$^{-2}$. This is plotted as a dashed line in Fig. \ref{fig1}. The cusp  near $V_b \simeq V_g \simeq 0$ and the absolute values for $V_b$ and $ V_g$ at the threshold are sensitive to the depth of the overetch and to the width which largely influence the coupling to the substrate. 
 
We start with one single donor located at $(x=0, y=15, z=3)$\,nm. Donor positions (see Fig. \ref{fig1} top panels) are given by $x$ (along S-D, $x$=0 means center, $|x|\ge$7.5\,nm are the extension regions), $y$ (transverse to S-D, $y=0$ means center, $y=\pm$25\,nm is the vertical edge of the nanowire), $z$ (vertical, $z$=0 means BOX interface, $z$=17\,nm is the top of the nanowire). This particular position is chosen such that it approximately corresponds to the experimental ionization curve for dopant A (see Fig. \ref{fig1} black line).

The ionization line is curved in ($V_g$, $V_b$), which would not be captured with a model assuming  constant capacitive couplings between the donor and the electrodes. The curvature means that the coupling to the gate and substrate are changing with $V_b$ and $ V_g$, as  a result of the ionization of donors in the extensions  and of the accumulation of surface carriers.  If we had chosen fully metallic, constant potential source and drain, the ionization line for an isolated donor would have been a straight line in the ($V_g$, $V_b$) plane.
Taking into account this complex electrostatic environment is absolutely necessary to describe qualitatively and quantitatively the ionization lines of the donors in the channel.

In particular the ionization lines become quasi-vertical, \textit{i.e.} less dependent on $V_g$, when $V_b$ is decreased towards negative values. The donor's ionization thus occurs at higher $V_g$ values where the conduction channel is set in the extensions of S-D. This screens the gate potential at the bottom of the channel, therefore on the donor, which becomes insensitive to $V_g$. 

\begin{figure}[t]
\begin{center}
\includegraphics[width=8.5cm]{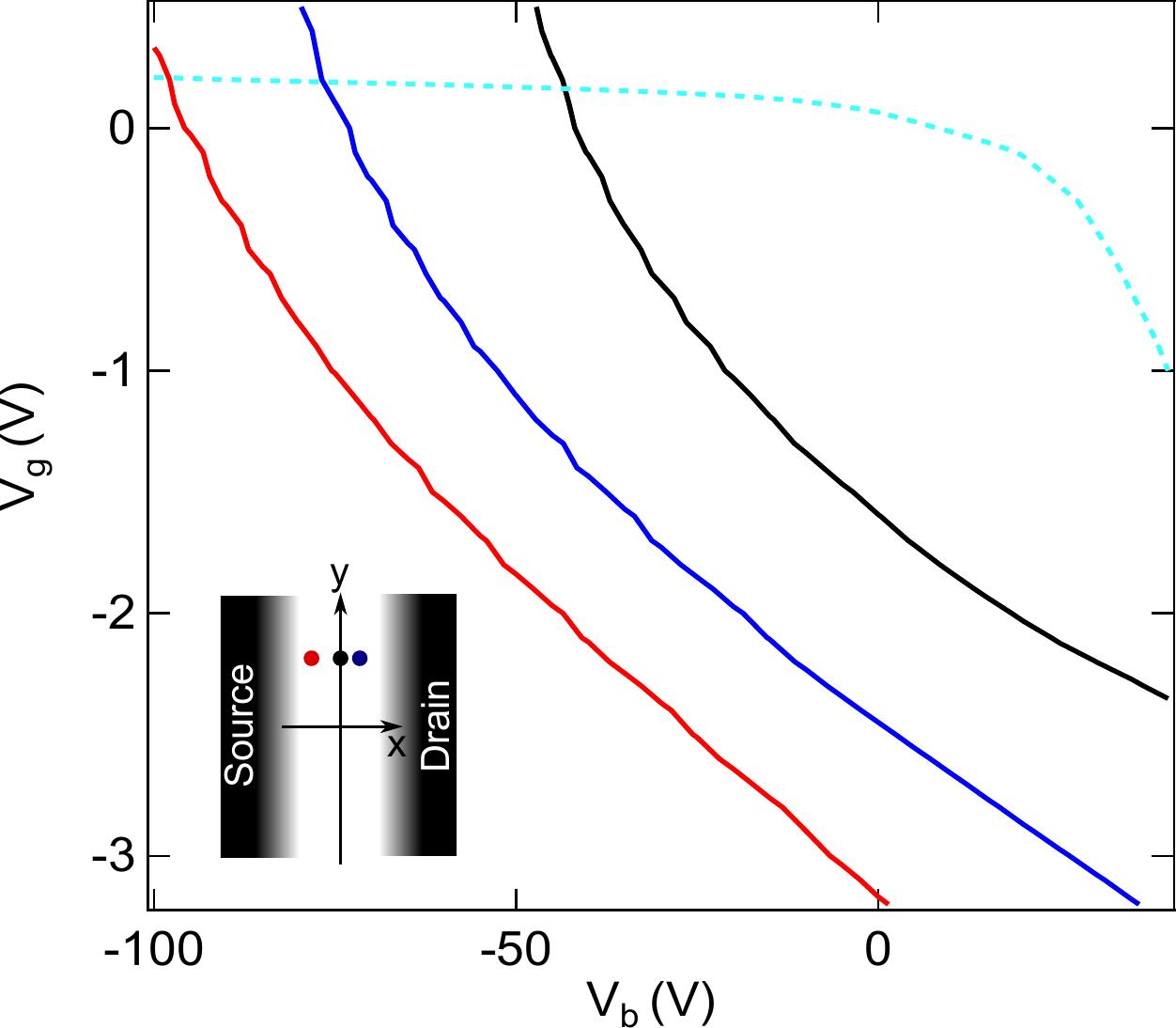}
\caption{(color online). Simulation of ionization \textit{vs.} ($V_b$, $V_g$) for a configuration with 3 donors in the channel at position (x, y=15\,nm, z=3\,nm). The donors differ by their x position (\textit{i.e.} along the S-D separation), as sketched in the inset: x=0 (resp. 3, -5) for the donor represented in black (resp. blue, red). The donor represented in black is at the same position that the simulated donor represented in black in Fig \ref{fig1}.} 
\label{fig2}
\end{center}
\end{figure}

\begin{figure}[t]
\begin{center}
\includegraphics[width=8.5cm]{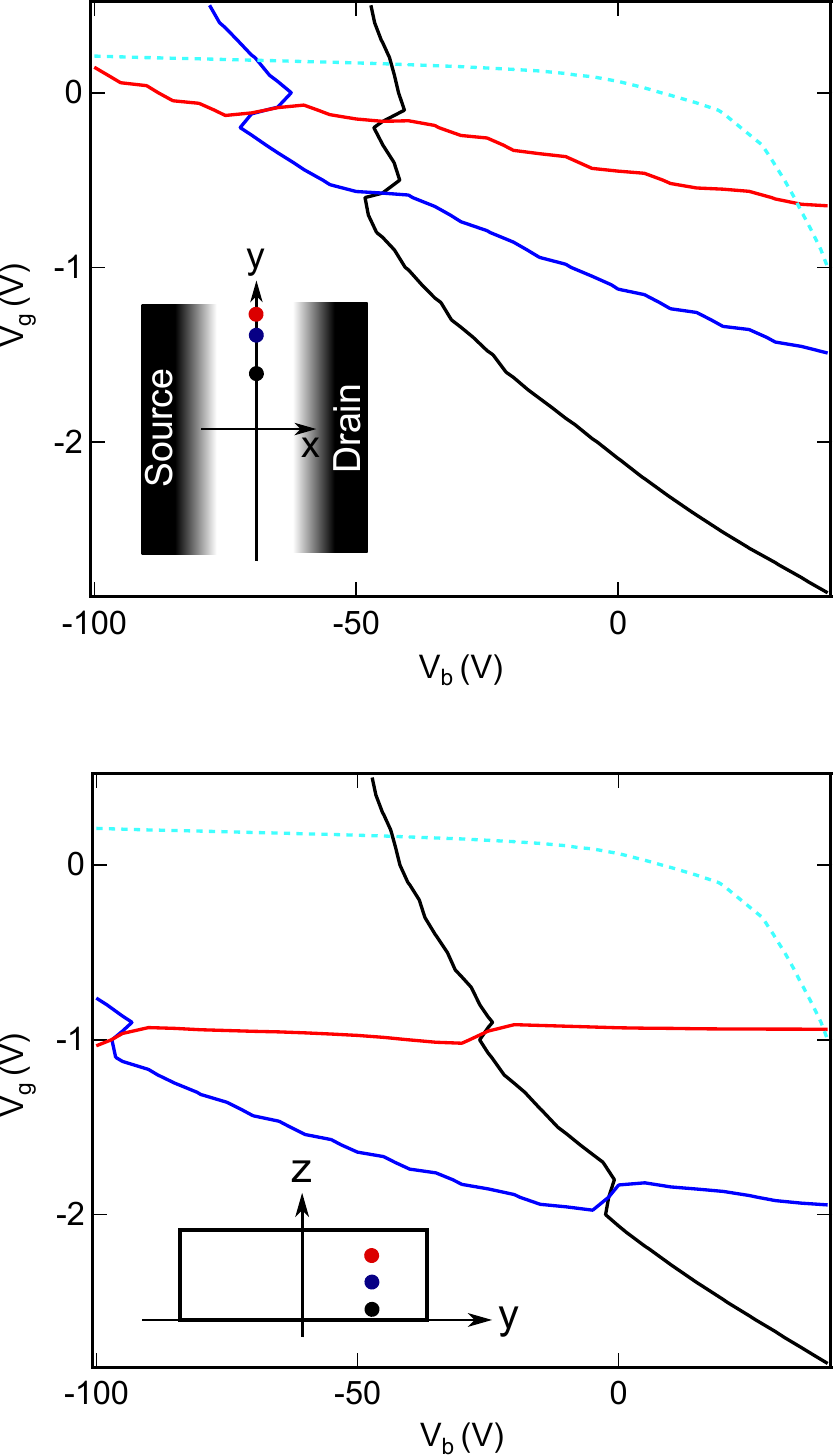}
\caption{(color online). Two simulations of ionization \textit{vs.} ($V_b$, $V_g$) for a configuration with 3 donors in the channel.  Donors differ by their y coordinates (top panel, x=0, y=15, 20, 23 \,nm, z=3\,nm) or z coordinates (bottom panel, x=0, y=15\,nm, z=3, 7.5, 12\,nm) as sketched in the insets.
The red donor (0, 23\,nm, 3\,nm) in the top panel is similar to the donor C-C' in Fig. \ref{fig1}}
\label{fig3}
\end{center}
\end{figure}

Then we introduce in the simulation two other donors whose positions differ either in $x$, $y$ or $z$ (see Fig. \ref{fig2} and \ref{fig3}). 
A change in $x$ (with  constant $y$, $z$, (see fig. \ref{fig2})) produces 3 almost parallel ionization lines. This is because the lever arm parameters change ($\alpha_g= {\delta \phi \over \delta V_g }$) and ($\alpha_b= {\delta \phi \over \delta V_b }$) where $\phi$ is the potential at the position of the donor. A donor centered in the channel has larger lever arm parameters (which means a better electrostatic control by the gate and substrate voltage) than a donor located closer to the S-D. In other words, there is a  significant electric field along $x$ in our structure when finite $ V_g$ and $ V_b$ are applied. The parallelism between the three ionization lines suggests that the ratio of ${\alpha_g \over \alpha_{b}}$ is barely affected. As a result donors close to the S-D ($x\simeq 5$) are first charged ($As^+$ to $As^0$) at very negative $V_g$ values. Donors more centered in the channel ($x \simeq 0$ or 3) are still ionized at  this value of $V_g$ (their energy states are above the Fermi energy in the S-D thanks to the potential gradient of potential along $x$).

On the contrary a change in $y$ or in $z$ modifies  the distance between the donor and the gates, which influences very much the curvature of the ionization lines,  \textit{i.e.} ${\alpha_g \over \alpha_b}$ (see Fig. \ref{fig3}):
donors located at the bottom center of the channel (near the BOX) are charged first at $V_b \gg $0\,V and $V_g \ll $0\,V. 
Donors located closer to the front gate (large $y$ or large $z$) are charged at larger  $V_g$ and are less sensitive to the backgate voltage, like the C-C' ionization lines in Fig. \ref{fig1}.
Therefore a measurement of the ionization lines as function of $V_g$ and $V_b$ allows to evaluate if a donor is close to the front gate or to the BOX.

It also allows to estimate the distance between donors. In fact ionization lines for one donor near the box and one donor near the top gate intercept each other at some point in ($V_b$, $V_g$).
At the intersection points, the lines anticross due to the repulsive Coulomb interaction between electrons on the two donors.
The shift in $V_{g}$ for an ionization line $i$ is given by ${(\alpha_{i,g})}^{-1} \times U_{ij}$ where $U_{ij}$ is the screened  Coulomb interaction  between electrons on donors $i$ and $j$. $U_{ij}$ is a sensitive function of the distance between donors as shown in Fig. \ref{fig3}, which brings some information about inter-donor coupling and their distance.
In particular we notice in Fig. \ref{fig3} that the ionization line for the donor (0, 15, 3), always represented in black, is shifted towards higher energy (\textit{i.e.} towards lower gate voltage) when the two other donors get ionized. 


With the help of these simulations, we can deduce the relative position  of the donors responsible for the six ionization lines -named A,A',B,B',C,C'- with respect to the FOX and the BOX (Fig. \ref{fig1}).
A striking observation in Fig. \ref{fig1} is that the A-A' then B-B' then C-C' ionizations lines  run approximately parallel to each other. 
 
Pairs of parallel lines can be due to i) 2 distant donors differing by their distance $x$ to the S-D electrodes -see Fig. \ref{fig2}-, or ii) the double occupation of a single donor ($As^+$/$As^0$ and $As^0$/$As^-$ ionization lines separated by the intra-donor charging energy). 
The intra-donor charging energy shifts the binding energy for the doubly occupied state $As^-$ closer to the conduction band. If the charging energy does not depend on the electric field, ionization lines are running parallel.
 
Because 3 pairs of lines running exactly parallel to each other will be very rare for a 6-donor configuration, the 3 pairs of lines are attributed to the double occupation of 3 separate arsenic donors.
The shape of the ionization lines A-A' and B-B' indicate two donors near the BOX and centered in the channel \textit{i.e.} small $y$ and $z$. 
By contrast lines C-C' are attributed to a donor close to the front gate (large $y$ or $z$, like the red lines in Fig.\ref{fig3}). We cannot take into account the double occupation problem in the simulation yet, as it would deserve to include electron-electron interactions beyond the mean-field treatment which was used. 
  
\begin{figure}[t]
\begin{center}
\includegraphics[width=8.5cm]{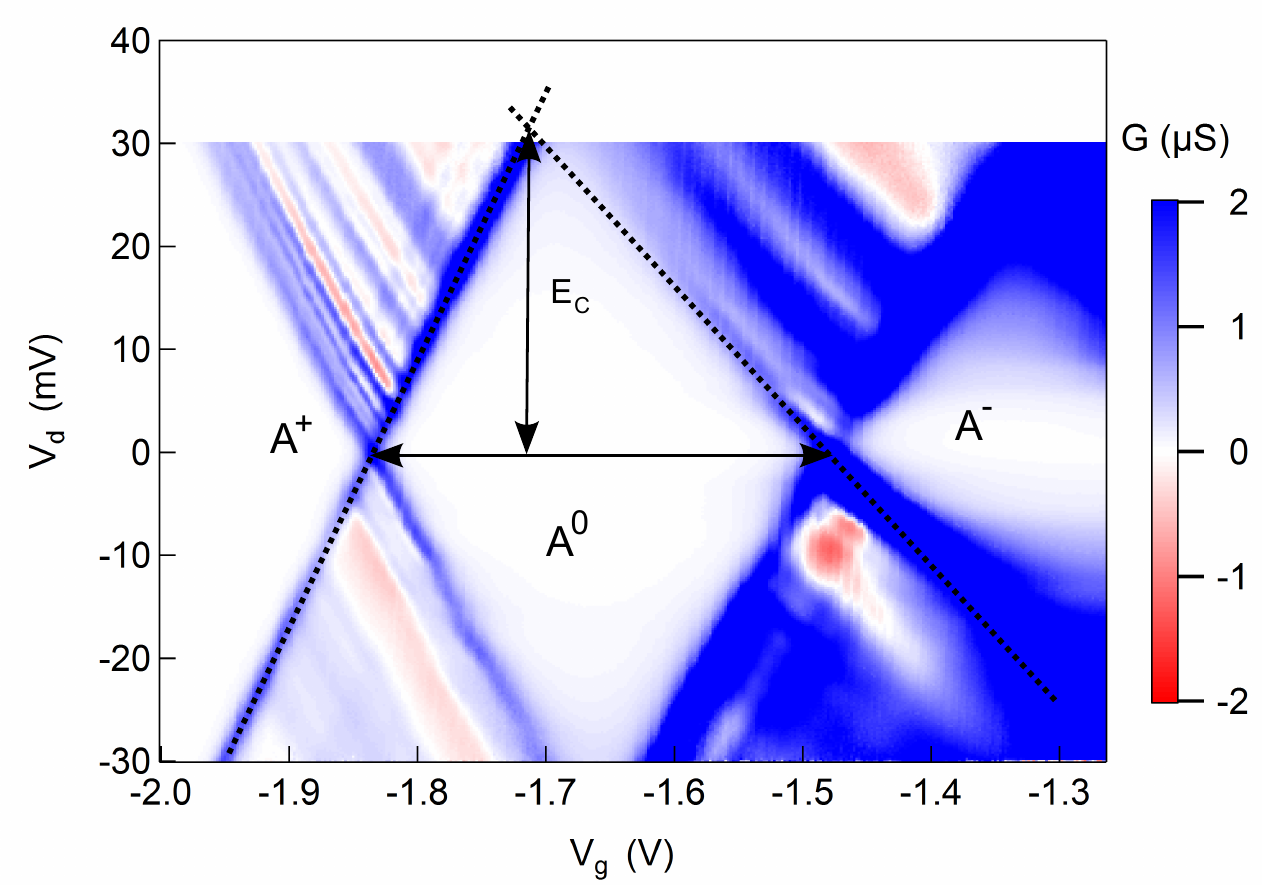}
\caption{(color online). Color plot of the S-D conductance at T=4.2\,K and $V_b$=10\,V. The two resonances are for ionization lines A and A', corresponding to the $As^+$/$As^0$ and $As^0$/$As^-$ transitions with a charging energy E$_c$. The lever arm factor is smaller for A' than for A as expected (see text). The  lines of differential conductance appearing at finite $V_d$ are due to local density of states fluctuation in the S-D\,\cite{Pierre2009}. They present approximately the same pattern for A and  A' but they are more blurred for A' due to higher tunneling rates.}
\label{fig4}
\end{center}
\end{figure}

Relatively small anticrossings are seen in the experiment. From the inset of fig. \ref{fig1} and from the measured value of the lever arm parameter (see later on, studies at finite $V_d$ as in fig. \ref{fig4} give $\alpha_{g} \simeq$ 0.1)  we deduce that the screened Coulomb interaction between donor A and C is  $U_{AC} \simeq $1.5\,meV,  which is also typical for the other anticrossings. This value is the bare Coulomb interaction  between two donors separated by 82\,nm, which is the maximal distance possible in our channel. More likely the bare Coulomb interaction is  screened by the S-D electrodes, which are distant by at most 5-10\,nm  from each donor, and by the gate, which is separated from the channel by 4\,nm of SiO$_2$. The anticrossings are smaller than simulated for donors separated by 9\,nm or less, indicating that the distance between donors A, B and C is larger than 10\,nm. 

The strong electric field in the channel combined with the small number of donors also favours the population of the $As^{-}$ state of a donor (A') rather than the ground $As^{0}$ state of another distant donor (B).
From studies at finite $V_d$ -as in Fig.~\ref{fig4}- we obtained  the energy separation $\simeq $ 50-60\,meV between the ground state for the two donors A and B ionized at $V_{b}$=10\,V.
This separation is due to the large electric field existing in the channel. In our simulation two donors with the same $E_i$ but slightly different $x$ (\textit{i.e.} distance to the S-D) are charged at very different gate voltages: at $V_{b}$=10\,V $\Delta V_{g} \simeq$=0.7\,V for $x=0$ and $x$=3\,nm (see Fig.  \ref{fig2}). Using a mean lever-arm factor of about 0.1 we estimate a field difference of 70\,meV in the S-D direction between $x$=0 and $x$=3\,nm. Therefore the separation between ionization lines A and B can be attributed to a variation in the $x$ coordinates of about $x_{A} - x_{B} \simeq 2-3 nm $.


Fig. \ref{fig4} provides a direct measurement of the lever-arm parameter $\alpha_{g}$ and of the energy separation between ionization lines A and A', attributed to the charging energy $E_c$. $E_c$ corresponds to the  value of $eV_{D}$ at the tip of the rhombus separating two ionization regions: $E_c \simeq $ 30\,meV for A-A' and 20\,meV for B-B'.
For C-C' we can only provide a lower bound $E_{c} \ge $30\,meV  for the charging energy because of the lack of contrast of the second resonance, which is too close from the threshold (not shown). The lever arm parameter is smaller for A' ($\alpha_{g}^{A'} $=0.08) than for A ($\alpha_{G}^{A} $=0.12), \textit{i.e.} $As^-$ is more strongly coupled to S-D than $As^0$. Two physical mechanisms account for this observation. First, the $As^-$ electronic orbital is  less bounded to the donor. Secondly, the ionization of $As^-$ occurs at higher $V_g$ where the S-D  are more extended towards the donor. These two mechanisms tend to increase $C_s$ and $C_d$ with respect to $C_g$ and thus result in a lower level-arm parameter for $As^-$ and a better donor orbital coupling to S-D.
One can also observe differential conductance lines appearing at finite $V_d$, parallel to edges of the diamond: these are due to local density of states fluctuations in the S-D\cite{Pierre2009}. Their patterns are approximately identical for A and A' but they are smoothed for A' due to higher tunneling rates. This supports the assumption that A and A' states are different ionization states associated to the same donor because they feel the same local environment in the S-D.

Several conclusions can be drawn from our observations: first the doubly charged state exists for the 3 donors. Second, the charging energy depends on the actual donor as the mesoscopic environment influences both the binding and the charging energies ($E_i$ and $E_c$)\,\cite{Diarra2007,Rahman2011}.

Third, the measured $E_c$ is much smaller than the ionization energy for As donors in bulk ($\simeq$ 53\,meV). The double occupied state is well separated from the conduction band threshold in the channel (see for instance the large separation between the A' line and the conduction band) , which means that the double occupied state of a donor is more stable in our nanostructure than in the bulk case.

Fourth, $E_c$ does not depend significantly on the electric field in the channel controlled by $V_b$, because pairs of ionisation lines run parallel to each other in a large range of ($V_b,V_g$). This is a new result because the electric field could not be varied on demand in previous experiments\,\cite{Rahman2011}.  

The stability of the double occupied state on shallow donors in the presence of an interface has been the subject of intense research \cite{Hao2009,Calderon2010,Hao2011}. Several differences are expected  with respect to the bulk case similar to the H$^-$ ion. The reduction of the charging energy can be due to the screening by a metallic gate electrode separated from the silicon by a very thin dielectric barrier \cite{Calderon2010,Hao2011,Verduijn2013}. The effect should however be small in our case where the gate oxide is 4\,nm thick. Moreover, the predicted binding energy for the doubly occupied state is  very small with or without the gate.
In all these previous works the S-D electrodes are neglected. In our devices the  donors are always stronger coupled to the S-D than to the gate (small lever arm parameter, short channel) and the screening by the source-drain electrodes \cite{Perrin2013} should be stronger and dominant compared to the screening by the front gate. 

The charging energy can also be reduced if the donor state is hybridized  with a  Si/SiO$_2$ surface state in the presence of a strong electric field \cite{Rahman2011}. 
In particular the calculated charging energy is found between 20\,meV and 30\, meV for donors 3 to 5\,nm from the interface in presence of a strong transverse electric field (of the order of 30\,meV/nm)\cite{Rahman2011} in good quantitative agreement with our results.
This simulation however does not include the screening by the gate and the S-D, which can further decrease the charging energy.
 
According to ref. \cite{Rahman2011} the fact that $E_C$ does not depend on the electric field   indicates that the electric field is always large in our device- in agreement with our simulation-, such that our donors -close to the BOX or to the FOX- are strongly hybridized with the interfacial state.

The charging sequence of the hybrid state could be the following. An As ion located a few nanometers away from an interface in the bulk is always ionized at large electric fields. It creates a local positive potential which forms a donor-induced potential dip at the interface. This dip attracts an electron and forms the singly-occupied state $As^0$. It is important to note that the first electron does not fully screen the $As^+$ ion but rather forms a dielectric dipole with it (transverse to $x$). This dipole produces a local field which is larger than the screened central potential which would result from the singly occupied neutral state in absence of electric field. The doubly occupied state  $As^-$ could be stabilized in that situation even if it is hard to conclude definitely on this point as it involves complex correlation effects between the two electrons and their image charges at the interfaces.
This scenario may explain why the lines A and A' run parallel to the conduction band edge at large positive $V_b$, because the interface 2DEG and the Coulomb island induced by the ionized donor potential have exactly the same coupling to the substrate and to the front gate.
 Both effects -the shift of the electron from the donor under large transverse electric field (along $y$, $z$) and the screening by the S-D electrodes (along $x$)- can add to explain the reduction of the charging energy and the stabilization of the $As^-$ state. However, a full simulation of the two-electron problem is lacking for a more quantitative analysis.


In summary we have tuned independently the ionization state of 3 randomly implanted  As donors in a nanoscale silicon MOSFET channel at low temperature, by applying both a front gate and a substrate bias.
At low energy, below the onset of the surface channels the dominant contribution to the S-D current is due to resonant tunneling through the hybrid donor-surface  $As^0$ and $As^-$  states of 3 As donors. 
In our very small devices where high electric fields are applied, the donors are hybridized with surface states.
Because the channel is very short the highly doped S-D plays the dominant role in screening the Coulomb interaction between distant donors. Combined with the transverse electric field this screening can also stabilize the double occupied state of a donor. The ionization of donors in the extension regions explains the evolution of the lever-arm parameter with the gate voltage as well as the shape of the ionization lines in ($V_g$, $V_b$).

Apart from the control of charges on donors by gates, the shape and position of the ionization lines in ($V_g$, $V_b$) and their evolution in $V_d$, which gives the respective couplings of donors to gate and S-D, can be used to perform a tomography of the donor's distribution in the channel~\cite{Khalafalla2007,Khalafalla2009,Mohiyaddin2013}. This tomography has not been presented here but our low temperature spectroscopy allows to measure in great details the random dopant fluctuations which is an well-known and major issue for microelectronics today.

\begin{acknowledgments} 
 We thank B. Sklenard and O. Cueto for extensive process simulation. The authors acknowledge financial support from the EC FP7 MINECC initiative under Project TOLOP No 318397 and the French ANR under Project SIMPSSON No 2010-Blan-1015.
\end{acknowledgments}

\newpage

%
%



\end{document}


\title{{\it Supplementary material} \\ --- \\Control of the ionization state of 3 single donor atoms in silicon.}

\author{B. Voisin}
\affiliation{SPSMS, UMR-E CEA / UJF-Grenoble 1, INAC, 17 rue des Martyrs, 38054 Grenoble, France}
\author{M. Cobian}
\affiliation{SPMM, UMR-E CEA / UJF-Grenoble 1, INAC, 17 rue des Martyrs, 38054 Grenoble, France}
\author{X. Jehl}
\affiliation{SPSMS, UMR-E CEA / UJF-Grenoble 1, INAC, 17 rue des Martyrs, 38054 Grenoble, France}
\author{M. Vinet}
\affiliation{CEA, LETI, MINATEC Campus, 17 rue des Martyrs, 38054 Grenoble, France}
\author{Y.-M. Niquet}
\affiliation{SPMM, UMR-E CEA / UJF-Grenoble 1, INAC, 17 rue des Martyrs, 38054 Grenoble, France}
\author{C. Delerue}
\affiliation{IEMN, 41 boulevard Vauban, 59046 Lille, France}
\author{S. de Franceschi}
\affiliation{SPSMS, UMR-E CEA / UJF-Grenoble 1, INAC, 17 rue des Martyrs, 38054 Grenoble, France}
\author{M. Sanquer}
\email[]{marc.sanquer@cea.fr}
\affiliation{SPSMS, UMR-E CEA / UJF-Grenoble 1, INAC, 17 rue des Martyrs, 38054 Grenoble, France}

\maketitle

To model the ionization lines, we treat the few donors in the channel as interacting point charges, and the donors in the highly-doped source and drain extensions as a continuum.

\begin{figure}[!t]
\centering
\includegraphics[width=\columnwidth]{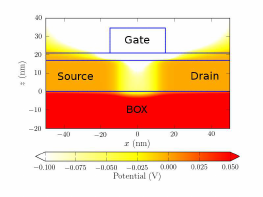}
\caption{Potential in the empty channel at bias point $V_g=-0.1$ V and $V_b=10$ V.}
\label{fig_sm}
\end{figure}

We first compute the potential $V(\vec{r})$ in the {\it empty} channel as a function of the front gate voltage $V_g$ and back gate voltage $V_b$, at zero source-drain bias. At that point we only account, therefore, for the donors in the highly doped source and drain extensions. We solve Poisson's equation self-consistently:
\begin{equation}
\nabla\kappa(\vec{r})\nabla V(\vec{r})=-4\pi\rho(\vec{r})\,,
\end{equation}
where $\kappa(\vec{r})$ is the dielectric constant at point $\vec{r}$ and $\rho(\vec{r})$ is the charge density. The latter reads:
\begin{equation}
\rho(\vec{r})=e[N_d^+(\vec{r})-n(\vec{r})]\,,
\end{equation}
where $N_d^+(\vec{r})$ is the density of ionized donors and $n(\vec{r})$ the density of electrons in the conduction band. We use the expression of Refs. \onlinecite{Altermatt2006a} and \onlinecite{Altermatt2006b} for $N_d^+(\vec{r})$, which is valid at high donor concentration $N_d$ and down to low temperature $T$:
\begin{multline}
N_d^+(\vec{r})/N_d(\vec{r})=\\
1-\frac{b}{1+\frac{1}{2}\exp[\beta(E_c-E_{\rm dop}-eV(\vec{r})-\mu)]}\,,
\end{multline}
where $E_c$ is the conduction band edge energy, $E_{\rm dop}$ is the donor binding energy, $\mu$ is the chemical potential, and $\beta=1/(k_{B}T)$, with $T$ the temperature. $E_{\rm dop}$ and $b$ are dependent on $N_d(\vec{r})$ as detailed in Ref. \onlinecite{Altermatt2006b}. As for $n(\vec{r})$, we make a semi-classical approximation:
\begin{equation}
n(\vec{r})=N_c F_{1/2}[\beta(E_c-eV(\vec{r})-\mu)]\,,
\end{equation}
where $F_{1/2}$ is the Fermi integral. Although crude, this simple model shall give a fair account of the screening by the quasi-metallic source and drain extensions. The above non-linear set of equations is solved with a Newton-Raphson method on a finite differences grid. For practical reasons (in particular, convergence), the potential landscape is computed at $T=30$ K. As an illustration, the potentiel at $V_g=-0.1$ V and $V_b=10$ V is plotted in Fig. \ref{fig_sm} of this supplementary mater.

Once the potential landscape has been computed as a function of $V_b$ and $V_g$, we add $N$ arsenic donors in the channel at positions $\{\vec{r}_i\}$. We assume that they behave as bulk-like donors, and track their bound-state energy levels 
\begin{equation}
E_{1s}(V_b, V_g, \vec{r}_i)=E_c-E_b-eV(\vec{r}_i)\,,
\end{equation}
where $E_b$ is the binding energy ($E_b=53$ meV for bulk arsenic donors). We also compute the Coulomb interactions $U_{ij}$ between these donors, as the interactions between point charges on our finite differences grid. We then write down the total energy of the system of donors for a given set of occupation numbers $\{n_i\}$:
\begin{multline}
E(\{n_i\})=\sum_{i=1}^N n_i[E_c-E_b-eV(\vec{r}_i)-\mu]\\
+\sum_{i=1}^N \sum_{j>i}^N (1-n_i)(1-n_j)U_{ij}\,,
\end{multline}
with $n_i=0$ for a ionized donor and $n_i=1$ for a neutral donor. The average occupation of a given donor finally reads:
\begin{equation}
\langle n_j\rangle=\frac{1}{Z}\sum_{\{n_i\}} n_j \exp[-E(\{n_i\})/kT]\,,
\end{equation}
where:
\begin{equation}
Z=\sum_{\{n_i\}} \exp[-E(\{n_i\})/kT]
\end{equation}
is the partition function. This model assumes that the donors have a bulk-like behavior, which might not be true in the possibly strong electric field that can be present in the device (see, e.g., the discussion about the hybridization with the channel at the end of the paper). It shall, however reproduce the main features and trends of the donor population.